\documentclass[sigplan,10pt]{acmart}
\AtBeginDocument{%
  }

\acmYear{2025}
\copyrightyear{2025}
\setcopyright{cc}
\setcctype[4.0]{by}
\acmConference[PLOS '25]{13th Workshop on Programming Languages and Operating Systems}{October 13--16, 2025}{Seoul, Republic of Korea}
\acmBooktitle{13th Workshop on Programming Languages and Operating Systems (PLOS '25), October 13--16, 2025, Seoul, Republic of Korea}
\acmDOI{10.1145/3764860.3768337}
\acmISBN{979-8-4007-2225-7/25/10}

\usepackage{booktabs}
\usepackage{multirow}
\usepackage{rotating}
 \usepackage{proof}
\usepackage{array}
\usepackage{iris}
\newcolumntype{O}[1]{>{\raggedright\arraybackslash}m{#1}}
\newcolumntype{P}[1]{>{\centering\arraybackslash}m{#1}}
\newcolumntype{Q}[1]{>{\raggedleft\arraybackslash}m{#1}}
\newcolumntype{S}{>{\centering\arraybackslash}m{2cm}}
\newcommand{\loc}{\textsf{Loc}}
\newcommand{\sepimp}{\mathrel{-\mkern-6mu*}}

\usepackage{amsthm}
\theoremstyle{definition}
\newtheorem{definition}{Definition}[section]
\begin{document}

\title{Modal Verification Patterns for Systems Software}

\author{Ismail Kuru}
\email{ik335@drexel.edu}
\orcid{0000-0002-5796-2150}
\affiliation{%
  \institution{Drexel University}
  \city{Philadelphia}
  \state{PA}
  \country{USA}
}
\author{Colin Gordon}
\email{csg63@drexel.edu}
\orcid{0000-0002-9012-4490}
\affiliation{%
  \institution{Drexel University}
  \city{Philadelphia}
  \state{PA}
  \country{USA}
}

\renewcommand{\shortauthors}{Kuru and Gordon}

\begin{abstract}
  Although they differ in the functionality they offer, low-level systems exhibit certain patterns of design and utilization of computing resources.
  In this paper we examine how modalities have emerged as a common structure in formal verification
  of low-level software, and explain how many recent examples naturally share 
  common structure in the relationship between the modalities and software features they are used
  to reason about.
  We explain how the concept of a \emph{resource context} (a class of system resources
  to reason about) naturally corresponds to families of modal operators indexed by
  system data, and how this naturally leads to using modal assertions to describe
  \emph{resource elements} (data in the relevant context).
\end{abstract}


%
%


\maketitle
\section{Introduction}
Although they differ in the functionality they offer, low-level systems exhibit certain patterns of design and utilization of computing resources.
Systems software, in general, interfaces with an underlying computing substrate such that any software system at any higher level in the software stack can (at least indirectly) utilize the resources of the machine.

The last layer of software before the hardware is naturally critical to the correctness of an overall system,
as essentially all software built on top of it assumes its correctness.
And because hardware is complex and highly diverse, the implementation of those lowest layers
of the software stack is typically intricate and naturally error-prone, despite how critical its correctness is.
Typically systems software has, as a primary focus, the task of \emph{abstracting} from hardware details to simplify 
the construction of higher layers of the stack.

Interacting with computing resources is the essential point which shapes the design of low-level systems software. 
Exploiting these design choices has been an important field of study.
This survey examines these common patterns in systems software, and their relationships
to modalities. We believe that certain properties of modalities enable us to understand and 
address the verification challenges of these systems by tailoring specification and proof
to the same patterns used to design the systems.

\textit{Contributions} We argue that modal abstractions can be used to identify and abstract system verification challenges. We justify our perspective by discussing prior systems that have successfully used modalities for system verification, arguing that they fit into the verification design pattern we articulate, and explaining how this approach might apply to other systems' verification challenges.
\begin{itemize}
\item \textbf{Identifying System Verification Challenges}: We start with identifying common patterns in system verification: \emph{virtualization, sharing, and translation}.
\item \textbf{Introducing Modal \emph{Resource Contexts}}: Then we discuss the concept of \emph{resource} which has already been an essential concept in the design of systems. Inspired by the concept of resource in systems, we define what a resource and its context are in the modal abstractions. For example, the \emph{transaction} abstraction, which behaves like a container for in-memory dirty disk pages, can be considered a resource context for any dirty in-memory disk page resource associated with it.
\item \textbf{Introducing \emph{Nominals for System Resources}}:
Existing logics for systems software must frequently identify
individual resources within a given context ---
nominalization (in the sense of hybrid logic) enables identifying a resource in its context. 
For example, a filesystem transaction is a context of resources for in-memory updated disk blocks
which have not yet been saved to disk.
The transaction identifier is used to associate a transaction with a disk block to persist so that, 
in case of a crash while persisting updated disk-blocks, the filesystem can rollback the 
\emph{already persisted} disk-blocks of the transaction, and restore the previous consistent disk state. 
We discuss other examples to contextualize the existing systems surveyed.
\item \textbf{Taxonomy of Current Modal Approaches in System Verification}: Based on these concepts, we summarize contemporary verification efforts using modal abstractions. We choose them from different domains, for example, reasoning about weak memory models and storage persistence ~\cite{tejthesis,perennialgit,tejperennial19,larsnextgen25,fsl,fsl++,derekrustbelt20,kuru2025modalabstractionsvirtualizingmemory,amalreal2024,gordon2019modal} and using different programming logics such as Iris ~\cite{jung2018iris} and Dafny ~\cite{leino2010dafny} because we would like to justify that our \emph{definitions} are neither domain nor programming logic dependent. 
\end{itemize}

\section{Background on Modal Logics for Programs, and Low-Level Systems Software}

This section briefly recalls related concepts from modal logic of particular relevance
to our taxonomy (Section \ref{sec:modalbackground}),
and enough general background on the systems software concepts organized by our taxonomy
for the connections to modalities to be clear (Section \ref{sec:background}).

\subsection{Hybrid Logic, Dynamic Logic, and Nominals in Program Logics}
\label{sec:modalbackground}
\paragraph{Dynamic Logics as Program Logics}
We believe that for the kinds of informal reasoning and the sorts of data structures discussed in the previous section, ideas from modal logic are a promising approach to formalization and verification.
Broadly speaking modal logics incorporate \emph{modal operators}, which take as arguments a proposition expected to be true in another time~\cite{pnueli1977temporal}, place~\cite{gordon2019modal,goranko1996hierarchies,areces2001hybrid,gargov1993modal,murphy2004symmetric}, circumstance or point-of-view~\cite{hintikka1962knowledge,halpern1985guide}, and result in a proposition true in the \emph{current} time, place, or circumstance at which the truth of the modal operator is being evaluated. Classic examples include modal necessity $\square P$ describing that $P$ is \emph{necessarily} true, $\mathsf{G}(P)$ meaning $P$ is true \emph{globally} (i.e., forever from this time onwards in temporal logics), or $K_i(P)$ describing that a particular participant $i$ \emph{knows} that $P$ is true. The latter is an example of \emph{multimodal} logic, where there is an family of modalities (modal operators) parameterized by some dimension of interest (there, participants).

\emph{Dynamic logic}~\cite{pratt1976semantical}, a (multi)modal logic variant of weakest preconditions~\cite{dijkstra-75,hahnle2022dijkstra,ahrendt2025many} 
which works with modalities of the form $[p](P)$, which states that \emph{in the current program state}, \emph{if} program $p$ is run then afterwards $P$ will hold (modulo non-termination).
\paragraph{Hybrid Logic} Hybrid logics~\cite{blackburn1995hybrid,goranko1996hierarchies,gargov1993modal,areces2001hybrid,brauner2010hybrid} are a class of modal logics that add two primary new concepts to a logic. Nominals $\ell \in \textsf{Loc}$ uniquely identify points in a model (e.g., a particular state in the state space the logic is used to reason about), so there is exactly one point in the model where a nominal $\loc$ (as a formula) is true. Satisfaction operators are modal operators indexed by nominals, which enable claims about the truth of another proposition at the point in the model identified by a chosen nominal. Traditionally $@_{\ell}\varphi$ asserts that $\varphi$ is true in the state identified by $\ell$. Conceptually, given our context, we might like $l$ to indicate truth in a particular data structure $\ell$.  However, most prior work on hybrid logics works in classical logics rather than a substructural setting like separation logic. 
Newer work~\cite{tejthesis,perennialgit,tejperennial19,larsnextgen25,fsl,fsl++,derekrustbelt20,kuru2025modalabstractionsvirtualizingmemory,amalreal2024}:w
 surveyed here adapts this idea to deal with different structures specified in
separation logic.
\paragraph{Nominalization}
Some classes of assertions benefit specifically from \emph{naming} the explicit conditions where they are true (as opposed to simply requiring them to be true \emph{somewhere} or \emph{everywhere} as in the most classic modal operators).
Because these named conditions strongly resemble the nominals of hybrid logic,
we refer to the general idea of naming circumstances explicitly as \emph{nominalization},
even though not all of the examples we discuss are actually hybrid logics.

An example use of state naming explicitly in the assertion appears in program logics 
with usage protocols (e.g. state transition systems resembling typestate~\cite{strom1986typestate,garcia2014foundations}, which ensure
the operations applied to some data occur in a specified order, and that invariants for various conceptual states of the data
--- such as a file being open or closed --- are respected). 
Protocol assertions are \emph{annotated with the name of the last (abstract) state at which the protocol is ensured}.
A classic application of this idea to systems is 
Halpern et al.'s work adapting modal logics of
knowledge to deal with distributed systems~\cite{halpern1990knowledge,halpern1989modelling,halpern2017reasoning}.
In most of that line of work, $\mathcal{K}_a(P)$ indicates that the node $a$ in the system \emph{knows} or \emph{possesses knowledge of} $P$
(for example, a Raft node may ``know'' a lower bound on the commit index).
Alternatively, a modality $@_i(P)$ may represent that $P$ is true of/at the specific node $i$~\cite{gordon2019modal} (e.g., that node $i$ has physically stored a certain
piece of data to reliable storage).
These permit capturing specific concepts relevant to the correctness (and reasoning about correctness) of a certain class of systems,
involving facts about specific named entities in the system.
\\

In each of these cases, the fact that these facts are described using a modality $M$ with the core modal property that if $P \Rightarrow Q$, then $M(P) \Rightarrow M(Q)$. This means, for example, that if a
process $p$ knows that the commit index is greater than 5 (e.g., $\mathcal{K}_p(\mathsf{commitIndex} > 5)$)
no extra work is required to conclude that the node knows it is greater than 3 (i.e., that $\mathcal{K}_p(\mathsf{commitIndex} > 3)$),
because this follows from standard properties of modal operators as described above.
In contrast, if verification instead used a custom assertion \[\mathsf{minCommitIndex}(5) \] to represent the former knowledge,
one would need to separately provide custom reasoning to conclude \[\mathsf{minCommitIndex}(3)\]
As our survey shows, the value of the modal view (which grants both intuitive specifications and natural
reasoning principles) is becoming apparent in systems software verification.

\subsection{Virtualization in Systems Software}
\label{sec:background}
One of the most common forms of abstraction provided by systems software is \emph{virtualization},
which abstracts the relationship between conceptual and physical computing resources.
Upper-levels of the software stack work with these conceptual --- \emph{virtual} ---
resources, while the lower levels of systems software deal with the translation
of requests expressed in terms of virtual resources into operations on physical resources.

Operating System (OS) kernels virtualize memory locations and quantity (via virtual memory and paging~\cite{Denning1970VM}).
Distributed language runtimes may virtualize addresses, even when processes may migrate across machines~\cite{jul1988fine}.
Filesystems virtualize locations on disk~\cite{rodeh2013btrfs,hitz1994file,Rosenblum1992LFS,bonwick2003zettabyte}.
Programs built on top of the corresponding systems software layer work logically at the level of these virtualized resources,
and it makes sense to specify the systems software directly in terms of those abstractions.

\paragraph{Access by Translation}
Accessing virtualized resources via translation is a common way to virtualize notions of location (e.g.,
virtual memory addresses, inodes or object IDs instead of disk addresses~\cite{bonwick2003zettabyte,hitz1994file,Rosenblum1992LFS,rodeh2013btrfs}).
B-trees, page tables, and related structures both behave like maps, when the corresponding physical resources exist as such
just in a different location.
Control over the lookup process (e.g, handling the case of a missing translation entry) allows for additional flexibility,
such as filling holes in sparse files, or demand paging (both from disk or lazily populating anonymous initially-zeroed mappings).
Although the realization of these maps may differ from system to system based on the context (and sometimes hardware details),
they are semantically --- logically --- partial maps, worth treating as such in verification.

\subsubsection{Virtualization of Memory Locations}
\label{sec:backgroundonmachinemodel}
\begin{figure}
    \includegraphics[width=\columnwidth]{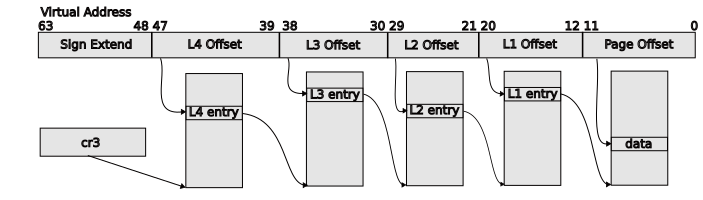}
    \caption{x86-64 page table lookups.}
    \label{fig:pagetables}
\end{figure}

A typical general purpose computer virtualizes  memory resources in RAM.
A program asking for a memory unit from an operating system kernel is served with a memory address that is \emph{virtualized}
such that it may be \emph{mapped to} a physical RAM location or not. 
As our informal definition suggests, the common technique for \emph{virtualization} of memory locations in OS 
kernels relies on \emph{translating} (\emph{mapping}) virtual addresses into physical ones. 
The conceptual address translation map is implemented with processor's page table trees as shown in 
Figure \ref{fig:pagetables}. In this Figure, we see a typical 4 levels of page tables, a virtual address on the 
top indexing into the different levels of the page table tree (the larger rectangles along the bottom
are blocks of memory which serve as nodes of the tree along a particular lookup path). 
Fabrication of virtual addresses out of limited physical addresses is provided with entries 
in the page table tree where each points to a physical address aligned to 4KB (4096 byte) boundaries. 
An address translation requires traversal of a series of tables starting from level 4. 
The traversal ends in the level 1 page table with the final lookup to the actual page of physical memory holding the requested data, and the low-order 12 bits being used to index into this page.

A page table tree is referred to as a (virtual) \emph{address space}, as it presents an abstraction
of how different memory addresses could be related in space.
This functionality is used by operating systems to present fictional collocation or separation of resources:
a range of logically (virtually) contiguous memory addresses may be translated to a range of physically disparate locations
in actual memory.
The address space then acts as a kind of container for virtualized memory resources tied to that container.

A typical OS manages multiple address spaces at once.
Each program is associated with a unique page table tree, which is stored in a specific hardware register (\texttt{cr3} in Figure \ref{fig:pagetables}). 
Using different mappings, which map only disjoint portions of physical memory in the map's codomain, 
is how the OS ensures memory isolation between processes. 
\subsubsection{Virtualized Resources in Filesystems}
Another important computer resource that is \emph{virtualized} for access is disk blocks. 
Filesystems are software components that abstract disk access on behalf of normal programs
Similar to how VMMs using page table trees to fabricate more memory addresses than what actually exists, 
filesystems use indexing maps for fabricating more \emph{virtual disk page} addresses which are conventionally called \emph{disk pages},
to be distinguished from physical \emph{disk block} addresses.
When a regular program accesses a file, it requests a specific logical offset from the start of that file's contents.
As with virtual memory, the filesystem may map adjacent logical blocks of a single file to physically disparate parts of the disk (or even
to different disks), though this translation mapping is implemented entirely in software (as opposed to the hardware cooperation that exists
for memory virtualization).

Filesystems not only handle the address translation from a disk-page to a disk-block but also need to handle different
\emph{modes} of computing where to find disk resources: a disk block may live on a disk in stable storage,
or the most up-to-date version of a block may live only in memory (for the filesystem to write to disk at a later time, batched for efficiency).
To handle the consistency of the disk resources in different modes of execution, filesystems employ different policies when indexing the disk resource.
Which data-structure is going to be used to implement address translation from disk-pages to disk-blocks? 
What is going to be the \emph{update policy} to the indexing map implementing address translation, which defines the \emph{consistency policy} for the disk resources? 
Different filesystem implementations give different answers to these questions. 

While exceptions exist~\cite{bonwick2003zettabyte,rodeh2013btrfs,Rosenblum1992LFS,hitz1994file},
most filesystems will continue to store a given block of a given file in the same physical disk block for as long as the file exists.
These filesystems with in-place updates need to implement some form of \emph{journalling}~\cite{hagmann1987reimplementing,chutani1992episode} to ensure that a crash partway through an in-place update
can be rolled back using an undo journal that records the original contents before in-place modification (so original contents could be restored after a crash) 
or completed using a write-ahead journal that first writes new data in unused space before \emph{also} updating in place (so the journal entries could be re-applied after a crash). 
Because disk blocks associated with a file could now be in many states (on-disk, journaled but not updated in place, or both)
the specification of these systems is complex. And because some states are only possible after a system failure,
each line of filesystem code has two conceptual post-conditions: one referring to the successfully stored disk pages from a disk page container (e.g, a disk tree) , which can be uniquely identified by the last successful commit identifer (e.g., a transaction identifier or snapshot identifier), as the current view of the disk in the specification for continued execution, and one referring to the previous view of the disk in the specification for the stable state established in the event of a hardware failure immediately after that operation.
\subsubsection{Reference Counting Memory Reclamation}

It is safe to reclaim memory (for later reuse) when that memory allocation will no longer be used by the client program.
Since that property itself is undecidable, sound heuristics are used instead.
A common proxy for the program never accessing an allocation again is to check that there is no
access path from data a program can trivially access (e.g., local variables stored on the stack)
to a memory allocation --- that is, when the allocation is no longer \emph{reachable}.
In that case the program \emph{cannot} access the memory again, so the memory location can be reclaimed \emph{soundly} ---
i.e., without violating any access through a reference to it.

To reclaim the allocated memory location that is no longer being accessed, many systems use a well-known method called \emph{reference counting} in which 
each memory allocation includes not only its data but also a counter of how many pointers to that memory exist in the rest of the program.
Any time a new reference to the allocated memory is created, the  counter is incremented,
and every time an existing reference is destroyed (by overwriting or dropping) the counter is decremented.\footnote{This may be done
using library functions for a reference counting system, or the operations may be inserted by a compiler.}
These reference counts are a sound (but incomplete) way to approximate reachability: if a decrement of a reference count
decrements the count to 0, no other references to that memory exist in the program, and the memory can be reclaimed.

Reasoning about the correctness of such code requires reasoning about these reachability relations and reference counts, relative to specific
stack frames.
\subsubsection{Weak Memory Models}
Multicore and multiprocessor systems do not directly write every memory update back to memory (or even a coherent hardware cache)
individually, as this would be extremely slow. Instead, they effectively buffer reads and writes to and from memory
locally within each CPU core~\cite{owens2009better}\footnote{This is a layer below the better-known L1 cache, which is coherent across all CPUs in the system.}
so that operations can be partially reordered and batched to improve performance. For most code --- where the different CPUs are doing
independent work --- this is transparent to software. But fine-grained concurrent systems code which has multiple CPU cores
simultaneously coordinating access to shared memory must reason about these (effective) bufferings and reorderings, whose details vary
across CPU architectures.

Most of this coordination comes through reasoning about what \emph{will be true after local operations are published to memory},
or what \emph{will be true after remote operations become visible locally}~\cite{sieczkowski2015separation,dang2022compass,fsl}. Each of these flavors of contingent reasoning
depend on use of certain \emph{fence} or \emph{memory barrier} operations, which block execution of a CPU until either all incoming or outgoing
memory transactions are complete.
\subsubsection{Actor Model} An alternative approach to addressing the shared memory concurrency complication, i.e., data races, is to use actors ~\cite{actormodel,agha1986actors} by eliminating shared mutable state and enforcing updates on exclusive local mutable state via exchanging immutable messages among processes. This is a way of sequentializing the concurrent interleaved execution. Often correctness of one actor depends on \emph{knowledge} about other actors. In a typical consensus algorithm, for example, an operation is committed if a quorum of nodes reach an agreement on it.
 
Therefore, as expected, the specification of an actor in the current view would refer to the other actors' states and has to validate that the shared knowledge amongst actors is preserved ~\cite{gordon2019modal}.

\section{Contingency Decomposition of a System}
In this section, we argue the position that \emph{modalities}  
should be a go-to approach when specifying and verifying low-level systems code.
We explain how the concept of a \emph{resource context} helps guide the design of new modalities
for verification of systems code, and
we justify our perspective by discussing prior systems that have used modalities for systems verification
successfully, arguing that they fit into the verification design pattern we articulate,
and explaining how this approach might apply to other systems verification challenges.

To explain our ideas in the general systems understanding, we briefly recap some of the background and themes our ideas build on again, casting them in a certain way to bring out the relevance of our philosophy.
\subsection{Encoding Modalities in Program Logics}
Program logics which would like to enjoy modal propositions need to have an underlying model that allows encoding of the propositions. 
A Kripke model can be roughly seen as a triple of set of worlds (e.g., program states), a binary relation between worlds (e.g., transition relations between states) and a state interpretation function. For a modality $M$, $M(\phi)$ is true in one state/world $w$ if states related to $w$ by the binary relation satisfy $P$.
Whether this is all related states, or just one, or something else varies depending on what $M$ is meant to model, but the modality is always tied
to the relation of interest. For multimodal logics there are multiple relations.
For example, for Dynamic Logic, there is a modality for each program, and the relation for running program ``command'' (in old program logic parlance)
$C$ is the relationship between states before and after executing $C$.

\paragraph{Encoding Hoare Triples}  Following Pratt's original representation~\cite{pratt1976semantical}, we can encode a Hoare triple $\{P\}C\{Q\}$ (asserting that
if $P$ holds in some state, then executing $C$ will yield a state where $Q$ holds) as $P\rightarrow[C](Q)$. However, for this to hold for a specific programming language and assertion language, the modal representation requires a semantics in terms of an underlying model.

Conventionally, a Kripke model with its constituent must be picked. Gordon provides a Kripke model for an actor-based hybrid modal logic~\cite{gordon2019modal}, in which the transition relation is indexed by a particular command and a particular actor, relating pre- and post-states of a specific actor referenced by a nominal executing the command. Gordon encodes this hybrid logic as a library in the Dafny program logic~\cite{leino2010dafny} with the help of two state invariants and rely-guarantee assertions checking the stability of invariants. Wagner et al. take a similar approach to the design of their separation logic and utilize a Kripke model for reference-counted memory reasoning principles~\cite{amalreal2024}.

On the other hand, program logics such as \textsf{Iris}~\cite{jung2018iris} utilizes Kripke models for their separation algebras (a collection of \textsc{CMRA}s as an algebraic cousin of a BI-algebra~\cite{ohearn1999bunched}), modeling modalities for both heap extension and step indexing~\cite{nakano-00}. 
However in general, many modalities can be given proof-theoretic semantics directly within a logic. 
This is the case for \textsf{Iris}'s weakest precondition modality (which is defined within the \textsf{Iris} 
logic itself), as well as for a wide range of other modalities in the 
literature~\cite{restall2002introduction}. These modalities are typically multi-modal, 
with a form of roughly $M~\overline{ctxt}~\phi\triangleq\ldots\rightarrow \phi~\overline{ctxt}$,
such that the evaluation of the modality incorporates the $\overline{ctxt}$ arguments into the evaluation of $\phi$.

\paragraph{Shallow Modalities in Separation Logic}
\label{sec:modal_in_iris}
In a substructural setting~\cite{restall1993modalities,krebbers2017essence} the implication $\rightarrow$ 
is typically replaced with substructural implication $\sepimp$.
This definition also reveals a subtlety about the denotation of assertions $\phi$ --- they are effectively \emph{predicates} of some kind, functions from some possible world into some logical algebra such as a BI-algebra~\cite{ohearn1999bunched}. 
Taking this interpretation literally gives rise to a convenient way to embed new modalities in \textsf{Iris} 
in a way well-suited to many use cases. 
If $B$ is a type with a BI-algebra structure, then any space of predicate functions (e.g., $\mathsf{val}\rightarrow B$ as a simplification) also carries a BI-algebra structure, lifted from $B$, often called the \emph{pointwise lifting} of the algebra. A number of 
projects for systems software verification~\cite{tejthesis,perennialgit,tejperennial19,larsnextgen25,fsl,fsl++,derekrustbelt20,kuru2025modalabstractionsvirtualizingmemory,amalreal2024} now take advantage of this,
explicitly or implicitly lifting the BI-algebra structure from \textsf{Iris} itself to \textsf{Iris} predicates 
used as assertions within modalities. 
Connecting this to \textsf{Iris}'s typeclasses for interactive proofs~\cite{krebbers2018mosel,krebbers-17} then permits working within the embedded logic just as if working in \textsf{Iris}.

In addition, the modal definitions themselves may represent ownership of resources in a substructural setting~\cite{restall1993modalities,dovsen1992modal,kamide2002kripke}. 
This allows modalities to simultaneously represent ownership of certain resources used for interpreting the modality itself, as well as whatever resource ownership is implied by the modal argument.
\subsection{Decomposing a System into its Constituents \emph{Contingently}}
\label{sec:definitions}
\begin{table*}[t]\small
\centering
\caption{Modal Decomposition of Program-Logics.}
\label{table:decomposition}
\begin{tabular}{@{}lcccp{4cm}@{}}
  \toprule
{ Modality}& { Context} & { Elements}  &  { Nominalization} & { Context Steps} \\ \midrule
 Post-Crash~\cite{tejthesis,perennialgit,tejperennial19} &  $\Diamond \; P $  &  $  \ell \mapsto_{n}^{\overline{\gamma}} v $ &  Strong &   Crash Recovery    \\ 
 NextGen~\cite{larsnextgen25} &   $\overset{t}{\hookrightarrow} \; P$  &  Own (t(a))  &  Strong &  Determined Based on the Model$^{*}$   \\
 StackRegion$^{*}$  ~\cite{larsnextgen25} &  $\overset{ICut^{n}}{\hookrightarrow} \; P$      &  $\fbox{n} \; \ell \mapsto v$                                 &  Strong &  Alloc and Return to/from stack\\
 Actor~\cite{gordon2019modal}& @$_{\iota}$ P    &  Variable values &  Weak &  Send Message \\
 Memory-Fence~\cite{fsl,fsl++,derekrustbelt20} &  $\triangle_{\pi}$ and $\triangledown_{\pi}$      &  $\ell \mapsto v$                                 &  Weak &  Fence Acquire and Release  \\
 Address Space~\cite{kuru2025modalabstractionsvirtualizingmemory} &  [r]P     &   $\ell \mapsto v$ &  Weak                         &  Address Space Switch  \\ 
 Ref-Count~\cite{amalreal2024}& @$_{\ell}$ P    &  $\ell_1 \mapsto v$ &  Weak  &  Allocating, Dropping and Sharing a Reference \\ \bottomrule
\end{tabular}

{*The StackRegion Modality is an instance of NextGen (called the Independence Modality in \cite{larsnextgen25})}.
\end{table*}

Many existing program logics for system verifications have a common structure, which maps to modalities with a couple extra dimensions of design.
We summarize our discussion of these logics in Table \ref{table:decomposition}.
We discuss, based on examples, common aspects of how we intuitively think about correctness of systems code in many contexts,
articulate those pieces, and call out the commonalities across a range of systems.
\subsection{Resource}
Consider first the address space abstraction in an OS kernel. An address space of a process is a \textsf{container} of \textsf{virtual} addresses referencing data in memory. One would expect to have \textit{points-to} assertions from separation logic to specify \textit{ownership} of a memory reference pointing to some data. In the classic setting~\cite{Reynolds:2002:SLL:645683.664578}, such a points-to assertion $x\mapsto v$ asserts ownership of memory location $x$, and that the contents of that memory location are $v$.
But that ownership is relative to a specific
address space --- a specific container. We tend to think directly about what is true \emph{in an address space},
with the simplest piece being an association between a virtual address and the data it points to.
We call the simplest piece, in this and other examples, the \emph{resource element}:

\begin{definition}[Resource Elements]
The simplest atomic facts we want to work with in a particular setting, specific to that setting.
\end{definition}
By definition, the resource elements are specific to some limited domain or setting.
For example, knowing that a certain address points to a 32-bit signed integer representing 3 is knowledge restricted to a certain address space.
In general,  we call these domains that any resource element is tied to \emph{resource contexts}:
\begin{definition}[Resource Context]
A resource context is an abstraction, context, or container of resource elements of the same type, e.g., an address space of a process.
\end{definition}
We discuss a range of examples for each of these in turn.

Table \ref{table:decomposition} gives additional examples of systems and their corresponding resource elements and contexts where these elements reside, though none of the work in that table analyzes itself according to the structure we are giving.

Except for the Post-Crash-Modality, one can think of the resource contexts in the first column in Table \ref{table:decomposition} as containers for the corresponding resource elements in the second column.
\paragraph{Stack Regions} When reasoning about stack frame contents, the resource element would be a stack-memory points-to assertion ( $\fbox{n} \; \ell \mapsto v$)
indicating that a certain offset into stack region $n$ holds value $v$. 
The regions (frames) themselves are the contexts.

\paragraph{Virtual Memory} For virtual memory management, a \textit{virtual-points-to}, $\ell \mapsto v$, is an ownership assertion pairing a virtual address ($\ell$) with data ($v$) in an address space, which is a resource context for virtual address mappings, $\ell \mapsto v$, and uniquely identified with a root address. Logically, specifying an address space switch ~\cite{kuru2025modalabstractionsvirtualizingmemory} from the address space with the root address $r2$ to the one with the root address $r1$
\[
\{[r1] P \ast Q \} cr3 := r1 \{P \ast [r2]Q\}_{\textsf{cr3}}
\]
requires having the resource elements loaded into the memory 
\[P \triangleq \ell_1 \mapsto v ...\]
to be in the resource context of the address space modality with the root $r1$, and, once the new address space is loaded—i.e., cr3 is loaded with the other address space root address—the resource elements ($Q$) that were loaded in the previous view of the memory are now introduced to the resource context with the root $r2$.
\paragraph{Weak-Memory} When considering weak memory models, we consider address-value mappings of thread-local views in C11 memory model~\cite{lahav2017repairing}). Modalities $\Delta_\pi(P)$ and $\nabla_\pi(P)$ with $P$ represent the memory address-value mappings that held before or will hold after certain memory fence operations by thread $\pi$.
The elements are points-to assertions, each specific to views of memory after those fence operations.

\paragraph{Reference Counting} When dealing with reference counting APIs, we may care to specify reachability of memory nodes ($\ell \mapsto v$) in a certain context defined by a shared root address. A shared memory address $\ell$ can be the root of the graph ($@_{\ell}\; ( \ell_1 \mapsto v_1 \;\ast\; ... \; \ast \; \ell_j \mapsto v_j )$) that can be a container of memory nodes ($\ell_1 \mapsto v$) that are reachable from the root $\ell$. 
\paragraph{Actors} When specifying the consistency of an actor in the current view,  we need to refer to other actors' state: $@_{\iota}(P)$ asserts that $P$ is true in the actor referenced actor ID $\iota$. 
Assertions may mention local variables and state at the designated actor by name, so the interpretation of variables in assertions depends on which actor
the assertion applies to.
\paragraph{Post-Crash}
The resource element of Post-Crash Modality is not obvious in Table \ref{table:decomposition}, and needs a bit of explanation. Perennial, based on the Iris logic, has both disk-points-to assertions $d[\textsf{p}] \mapsto_{n} v$ (for a specific disk $d$) and in-memory points-to assertions $\ell \mapsto_{n}^{\overline{\gamma}}$. Perennial crash-recovery logic book-keeps resource names (can be thought of as logical variables) $\overline{\gamma}$ to identify 
which assertions (resource elements) remain valid after a crash --- these assertions
are only usable while the names in $\overline{\gamma}$ are valid, and a crash resets
them, discarding assumptions about volatile state. A subtlety of the notion of a resource context is that, unlike the earlier examples, the context does not need to be a literal data structure. It can instead be (various forms of) a set of executions, as in the Post-Crash and NextGen modalities.
The Post-Crash modality $\diamond$ expresses that the assertion $P$ will be true after a crash discards
all unstable storage (i.e., RAM).
This was the inspiration for the NextGen modality, which is in fact a framework for defining ``after-$t$'' modalities, where $t$ is an transformation of the global state.\footnote{The transformations are subject to some
technical constraints that are unrelated to our point here.}
\subsection{Nominals for Systems Resource Contexts}
Finally, another design point is the question of whether or not resource element assertions must explicitly track
their corresponding context, or if they implicitly pick up their context from where they are used.

\emph{Strong nominalization} is the case where resource elements must explicitly include
the identity of their intended context, while \emph{weak nominalization} occurs when the resource
elements implicitly pick up the relevant context from how they are used.
The first three modalities in Table \ref{table:decomposition} are strongly nominalized,
with the resource elements generally carrying identifiers of a specific modality usage.\footnote{The Post-Crash modality does not look like this in the presentation here; technically the definition of $\diamond$ quantifies over names $\overline{\gamma}$ internally, dealing with sets of possible contexts.}
The rest are weakly nominalized.

This choice trades off complexity against flexibility and scoping constraints.
Strongly nominalized elements track additional specifier/prover-visible book-keeping data.
But in exchange for carrying those identifiers of their context with them, 
strongly nominalized elements can be used together under any modality. For example, one can
use the StackRegion modality to talk about two different stack frames simultaneously for code
which accesses multiple stack frames: $\fbox{n} \; \ell \mapsto v \ast \fbox{n+1} \; \ell' \mapsto v'$.
Using a given strongly nominalized assertion element under different modalities for different
frames does not change its meaning. Likewise,  weak-memory specific modalities existentially quantify over other views, related to the ``current'' view (the one where
the current thread's assertions are evaluated), and evaluate a resource (a thread-local view of an address-value mapping) with respect to those other views. Given that views are identified by \emph{nominals}, referring to all the other views makes weak-memory modalities strongly nominalized.

By contrast, weakly nominalized elements are more concise, but then make talking about multiple 
contexts together marginally more complex: changing which modality an assertion is used with drastically changes its meaning. In the case of the Ref-Count modality, $@_\ell(\ell_1\mapsto v)$
says that $\ell$ points to a reference count wrapping $\ell_1$, while placing the $\ell_1\mapsto v$ under a jump modality for a different location entails talking about a different region of memory.

In general, use cases where code frequently manipulates small parts of multiple contexts together
should prefer strong nominalization, while use cases where usually larger portions of a single context
are at issue should prefer weak nominalization.

\section{Conclusion}
\label{sec:conclusion}
We have argued that there is a systematic pattern emerging in the use of modalities in the verification of low-level systems software,
and explained how a range of modern work fits into this pattern.
While most of our examples come from public verification work carried out in Iris,
we have one example of these ideas applying in Dafny.
We are applying these ideas in ongoing work using different modalities for different parts of a copy-on-write filesystem (e.g., for assertions true in different snapshots), and are observing more structured specifications as a result. 
So not only are the modalities observably present, but they are demonstrably useful
for conducting proofs of systems software.
We believe that these essential patterns in the designs of modal abstractions for systems verification constitute a fundamental
concept
when working out how to specify different kinds of systems code.
\bibliographystyle{ACM-Reference-Format}
\bibliography{software}


\begin{thebibliography}{50}


\ifx \showCODEN    \undefined \def \showCODEN     #1{\unskip}     \fi
\ifx \showISBNx    \undefined \def \showISBNx     #1{\unskip}     \fi
\ifx \showISBNxiii \undefined \def \showISBNxiii  #1{\unskip}     \fi
\ifx \showISSN     \undefined \def \showISSN      #1{\unskip}     \fi
\ifx \showLCCN     \undefined \def \showLCCN      #1{\unskip}     \fi
\ifx \shownote     \undefined \def \shownote      #1{#1}          \fi
\ifx \showarticletitle \undefined \def \showarticletitle #1{#1}   \fi
\ifx \showURL      \undefined \def \showURL       {\relax}        \fi
\providecommand\bibfield[2]{#2}
\providecommand\bibinfo[2]{#2}
\providecommand\natexlab[1]{#1}
\providecommand\showeprint[2][]{arXiv:#2}

\bibitem[Areces et~al\mbox{.}(2001)]%
        {areces2001hybrid}
\bibfield{author}{\bibinfo{person}{Carlos Areces}, \bibinfo{person}{Patrick
  Blackburn}, {and} \bibinfo{person}{Maarten Marx}.}
  \bibinfo{year}{2001}\natexlab{}.
\newblock \showarticletitle{Hybrid logics: Characterization, interpolation and
  complexity}.
\newblock \bibinfo{journal}{\emph{The Journal of Symbolic Logic}}
  \bibinfo{volume}{66}, \bibinfo{number}{3} (\bibinfo{year}{2001}),
  \bibinfo{pages}{977--1010}.
\newblock


\bibitem[Blackburn and Seligman(1995)]%
        {blackburn1995hybrid}
\bibfield{author}{\bibinfo{person}{Patrick Blackburn} {and}
  \bibinfo{person}{Jerry Seligman}.} \bibinfo{year}{1995}\natexlab{}.
\newblock \showarticletitle{Hybrid languages}.
\newblock \bibinfo{journal}{\emph{Journal of Logic, Language and Information}}
  \bibinfo{volume}{4}, \bibinfo{number}{3} (\bibinfo{year}{1995}),
  \bibinfo{pages}{251--272}.
\newblock


\bibitem[Bonwick et~al\mbox{.}(2003)]%
        {bonwick2003zettabyte}
\bibfield{author}{\bibinfo{person}{Jeff Bonwick}, \bibinfo{person}{Matt
  Ahrens}, \bibinfo{person}{Val Henson}, \bibinfo{person}{Mark Maybee}, {and}
  \bibinfo{person}{Mark Shellenbaum}.} \bibinfo{year}{2003}\natexlab{}.
\newblock \showarticletitle{{The Zettabyte File System}}. In
  \bibinfo{booktitle}{\emph{{Proc. of the 2nd Usenix Conference on File and
  Storage Technologies (USENIX FAST)}}}.
\newblock


\bibitem[Bra{\"u}ner(2010)]%
        {brauner2010hybrid}
\bibfield{author}{\bibinfo{person}{Torben Bra{\"u}ner}.}
  \bibinfo{year}{2010}\natexlab{}.
\newblock \bibinfo{booktitle}{\emph{Hybrid logic and its proof-theory}}.
  Vol.~\bibinfo{volume}{37}.
\newblock \bibinfo{publisher}{Springer Science \& Business Media}.
\newblock


\bibitem[Chajed(2022)]%
        {tejthesis}
\bibfield{author}{\bibinfo{person}{Tej Chajed}.}
  \bibinfo{year}{2022}\natexlab{}.
\newblock \emph{\bibinfo{title}{Verifying a concurrent, crash-safe file system
  with sequential reasoning}}.
\newblock Ph.D. Dissertation. \bibinfo{school}{Machetutes Institute of
  Technology}, \bibinfo{address}{Cambridge, MA}.
\newblock
\newblock
\shownote{Available at \url{ https://dspace.mit.edu/handle/1721.1/144578}}.


\bibitem[Chajed et~al\mbox{.}(2023)]%
        {perennialgit}
\bibfield{author}{\bibinfo{person}{Tej Chajed}, \bibinfo{person}{Joseph
  Tassarotti}, {and} \bibinfo{person}{Contributors}.}
  \bibinfo{year}{2023}\natexlab{}.
\newblock \bibinfo{title}{Post-crash modality in Perennial’s Coq
  Mechanization}.
\newblock
\urldef\tempurl%
\url{https://github.com/mit-pdos/perennial/blob/master/src/goose\_
  lang/crash\_modality.v}
\showURL{%
\tempurl}


\bibitem[Chajed et~al\mbox{.}(2019)]%
        {tejperennial19}
\bibfield{author}{\bibinfo{person}{Tej Chajed}, \bibinfo{person}{Joseph
  Tassarotti}, \bibinfo{person}{M.~Frans Kaashoek}, {and}
  \bibinfo{person}{Nickolai Zeldovich}.} \bibinfo{year}{2019}\natexlab{}.
\newblock \showarticletitle{Verifying concurrent, crash-safe systems with
  Perennial}. In \bibinfo{booktitle}{\emph{Proceedings of the 27th ACM
  Symposium on Operating Systems Principles}} (Huntsville, Ontario, Canada)
  \emph{(\bibinfo{series}{SOSP '19})}. \bibinfo{publisher}{Association for
  Computing Machinery}, \bibinfo{address}{New York, NY, USA},
  \bibinfo{pages}{243–258}.
\newblock
\showISBNx{9781450368735}
\href{https://doi.org/10.1145/3341301.3359632}{doi:\nolinkurl{10.1145/3341301.3359632}}


\bibitem[Chutani et~al\mbox{.}(1992)]%
        {chutani1992episode}
\bibfield{author}{\bibinfo{person}{Sailesh Chutani}, \bibinfo{person}{Owen~T
  Anderson}, \bibinfo{person}{Michael~L Kazar}, \bibinfo{person}{Bruce~W
  Leverett}, \bibinfo{person}{W~Anthony Mason}, {and} \bibinfo{person}{Robert~N
  Sidebotham}.} \bibinfo{year}{1992}\natexlab{}.
\newblock \showarticletitle{The Episode file system}. In
  \bibinfo{booktitle}{\emph{Proceedings of the USENIX Winter 1992 Technical
  Conference}}. San Fransisco, CA, USA, \bibinfo{pages}{43--60}.
\newblock


\bibitem[Dang et~al\mbox{.}(2019)]%
        {derekrustbelt20}
\bibfield{author}{\bibinfo{person}{Hoang-Hai Dang},
  \bibinfo{person}{Jacques-Henri Jourdan}, \bibinfo{person}{Jan-Oliver Kaiser},
  {and} \bibinfo{person}{Derek Dreyer}.} \bibinfo{year}{2019}\natexlab{}.
\newblock \showarticletitle{RustBelt meets relaxed memory}.
\newblock \bibinfo{journal}{\emph{Proc. ACM Program. Lang.}}
  \bibinfo{volume}{4}, \bibinfo{number}{POPL}, Article \bibinfo{articleno}{34}
  (\bibinfo{date}{Dec.} \bibinfo{year}{2019}), \bibinfo{numpages}{29}~pages.
\newblock
\href{https://doi.org/10.1145/3371102}{doi:\nolinkurl{10.1145/3371102}}


\bibitem[Dang et~al\mbox{.}(2022)]%
        {dang2022compass}
\bibfield{author}{\bibinfo{person}{Hoang-Hai Dang}, \bibinfo{person}{Jaehwang
  Jung}, \bibinfo{person}{Jaemin Choi}, \bibinfo{person}{Duc-Than Nguyen},
  \bibinfo{person}{William Mansky}, \bibinfo{person}{Jeehoon Kang}, {and}
  \bibinfo{person}{Derek Dreyer}.} \bibinfo{year}{2022}\natexlab{}.
\newblock \showarticletitle{Compass: strong and compositional library
  specifications in relaxed memory separation logic}. In
  \bibinfo{booktitle}{\emph{Proceedings of the 43rd ACM SIGPLAN International
  Conference on Programming Language Design and Implementation}}.
  \bibinfo{pages}{792--808}.
\newblock


\bibitem[Denning(1970)]%
        {Denning1970VM}
\bibfield{author}{\bibinfo{person}{Peter~J. Denning}.}
  \bibinfo{year}{1970}\natexlab{}.
\newblock \showarticletitle{{Virtual Memory}}.
\newblock \bibinfo{journal}{\emph{ACM Comput. Surv.}} \bibinfo{volume}{2},
  \bibinfo{number}{3} (\bibinfo{date}{Sept.} \bibinfo{year}{1970}),
  \bibinfo{pages}{153--189}.
\newblock
\showISSN{0360-0300}
\href{https://doi.org/10.1145/356571.356573}{doi:\nolinkurl{10.1145/356571.356573}}


\bibitem[Dijkstra(1975)]%
        {dijkstra-75}
\bibfield{author}{\bibinfo{person}{Edsger~W. Dijkstra}.}
  \bibinfo{year}{1975}\natexlab{}.
\newblock \showarticletitle{Guarded commands, nondeterminacy and formal
  derivation of programs}.
\newblock \bibinfo{journal}{\emph{Commun. ACM}} \bibinfo{volume}{18},
  \bibinfo{number}{8} (\bibinfo{year}{1975}), \bibinfo{pages}{453--457}.
\newblock
\urldef\tempurl%
\url{http://doi.acm.org/10.1145/360933.360975}
\showURL{%
\tempurl}


\bibitem[Doko and Vafeiadis(2016)]%
        {fsl}
\bibfield{author}{\bibinfo{person}{Marko Doko} {and} \bibinfo{person}{Viktor
  Vafeiadis}.} \bibinfo{year}{2016}\natexlab{}.
\newblock \showarticletitle{A Program Logic for {C11} Memory Fences}. In
  \bibinfo{booktitle}{\emph{Proceedings of the 17th International Conference on
  Verification, Model Checking, and Abstract Interpretation}} (St. Petersburg,
  FL, USA).
\newblock
\href{https://doi.org/10.1007/978-3-662-49122-5\_20}{doi:\nolinkurl{10.1007/978-3-662-49122-5\_20}}


\bibitem[Doko and Vafeiadis(2017)]%
        {fsl++}
\bibfield{author}{\bibinfo{person}{Marko Doko} {and} \bibinfo{person}{Viktor
  Vafeiadis}.} \bibinfo{year}{2017}\natexlab{}.
\newblock \showarticletitle{Tackling Real-Life Relaxed Concurrency with FSL++}.
  In \bibinfo{booktitle}{\emph{Programming Languages and Systems: 26th European
  Symposium on Programming, ESOP 2017}} (Uppsala, Sweden).
\newblock
\href{https://doi.org/10.1007/978-3-662-54434-1_17}{doi:\nolinkurl{10.1007/978-3-662-54434-1_17}}


\bibitem[Do{\v{s}}en(1992)]%
        {dovsen1992modal}
\bibfield{author}{\bibinfo{person}{Kosta Do{\v{s}}en}.}
  \bibinfo{year}{1992}\natexlab{}.
\newblock \showarticletitle{Modal translations in substructural logics}.
\newblock \bibinfo{journal}{\emph{Journal of Philosophical Logic}}
  \bibinfo{volume}{21}, \bibinfo{number}{3} (\bibinfo{year}{1992}),
  \bibinfo{pages}{283--336}.
\newblock


\bibitem[Garcia et~al\mbox{.}(2014)]%
        {garcia2014foundations}
\bibfield{author}{\bibinfo{person}{Ronald Garcia}, \bibinfo{person}{{\'E}ric
  Tanter}, \bibinfo{person}{Roger Wolff}, {and} \bibinfo{person}{Jonathan
  Aldrich}.} \bibinfo{year}{2014}\natexlab{}.
\newblock \showarticletitle{Foundations of typestate-oriented programming}.
\newblock \bibinfo{journal}{\emph{ACM Transactions on Programming Languages and
  Systems (TOPLAS)}} \bibinfo{volume}{36}, \bibinfo{number}{4}
  (\bibinfo{year}{2014}), \bibinfo{pages}{1--44}.
\newblock


\bibitem[Gargov and Goranko(1993)]%
        {gargov1993modal}
\bibfield{author}{\bibinfo{person}{George Gargov} {and}
  \bibinfo{person}{Valentin Goranko}.} \bibinfo{year}{1993}\natexlab{}.
\newblock \showarticletitle{Modal logic with names}.
\newblock \bibinfo{journal}{\emph{Journal of Philosophical Logic}}
  \bibinfo{volume}{22}, \bibinfo{number}{6} (\bibinfo{year}{1993}),
  \bibinfo{pages}{607--636}.
\newblock


\bibitem[Goranko(1996)]%
        {goranko1996hierarchies}
\bibfield{author}{\bibinfo{person}{Valentin Goranko}.}
  \bibinfo{year}{1996}\natexlab{}.
\newblock \showarticletitle{Hierarchies of modal and temporal logics with
  reference pointers}.
\newblock \bibinfo{journal}{\emph{Journal of Logic, Language and Information}}
  \bibinfo{volume}{5}, \bibinfo{number}{1} (\bibinfo{year}{1996}),
  \bibinfo{pages}{1--24}.
\newblock


\bibitem[Gordon(2019)]%
        {gordon2019modal}
\bibfield{author}{\bibinfo{person}{Colin~S Gordon}.}
  \bibinfo{year}{2019}\natexlab{}.
\newblock \showarticletitle{Modal assertions for actor correctness}. In
  \bibinfo{booktitle}{\emph{Proceedings of the 9th ACM SIGPLAN International
  Workshop on Programming Based on Actors, Agents, and Decentralized Control}}.
  \bibinfo{pages}{11--20}.
\newblock


\bibitem[Hagmann(1987)]%
        {hagmann1987reimplementing}
\bibfield{author}{\bibinfo{person}{Robert Hagmann}.}
  \bibinfo{year}{1987}\natexlab{}.
\newblock \showarticletitle{Reimplementing the Cedar file system using logging
  and group commit}. In \bibinfo{booktitle}{\emph{Proceedings of the eleventh
  ACM Symposium on Operating systems principles}}. \bibinfo{pages}{155--162}.
\newblock


\bibitem[Halpern(2017)]%
        {halpern2017reasoning}
\bibfield{author}{\bibinfo{person}{Joseph~Y Halpern}.}
  \bibinfo{year}{2017}\natexlab{}.
\newblock \bibinfo{booktitle}{\emph{Reasoning about uncertainty}}.
\newblock \bibinfo{publisher}{MIT press}.
\newblock


\bibitem[Halpern and Fagin(1989)]%
        {halpern1989modelling}
\bibfield{author}{\bibinfo{person}{Joseph~Y Halpern} {and}
  \bibinfo{person}{Ronald Fagin}.} \bibinfo{year}{1989}\natexlab{}.
\newblock \showarticletitle{Modelling knowledge and action in distributed
  systems}.
\newblock \bibinfo{journal}{\emph{Distributed computing}}  \bibinfo{volume}{3}
  (\bibinfo{year}{1989}), \bibinfo{pages}{159--177}.
\newblock


\bibitem[Halpern and Moses(1985)]%
        {halpern1985guide}
\bibfield{author}{\bibinfo{person}{Joseph~Y Halpern} {and}
  \bibinfo{person}{Yoram Moses}.} \bibinfo{year}{1985}\natexlab{}.
\newblock \showarticletitle{A guide to the modal logics of knowledge and
  belief}. In \bibinfo{booktitle}{\emph{Proceedings of the 9th international
  joint conference on Artificial intelligence-Volume 1}}. Morgan Kaufmann
  Publishers Inc., \bibinfo{pages}{480--490}.
\newblock


\bibitem[Halpern and Moses(1990)]%
        {halpern1990knowledge}
\bibfield{author}{\bibinfo{person}{Joseph~Y Halpern} {and}
  \bibinfo{person}{Yoram Moses}.} \bibinfo{year}{1990}\natexlab{}.
\newblock \showarticletitle{Knowledge and common knowledge in a distributed
  environment}.
\newblock \bibinfo{journal}{\emph{Journal of the ACM (JACM)}}
  \bibinfo{volume}{37}, \bibinfo{number}{3} (\bibinfo{year}{1990}),
  \bibinfo{pages}{549--587}.
\newblock


\bibitem[Hewitt et~al\mbox{.}(1973)]%
        {actormodel}
\bibfield{author}{\bibinfo{person}{Carl Hewitt}, \bibinfo{person}{Peter
  Bishop}, \bibinfo{person}{Irene Greif}, \bibinfo{person}{Brian Smith},
  \bibinfo{person}{Todd Matson}, {and} \bibinfo{person}{Richard Steiger}.}
  \bibinfo{year}{1973}\natexlab{}.
\newblock \showarticletitle{Actor induction and meta-evaluation}. In
  \bibinfo{booktitle}{\emph{Proceedings of the 1st Annual ACM SIGACT-SIGPLAN
  Symposium on Principles of Programming Languages}} (Boston, Massachusetts)
  \emph{(\bibinfo{series}{POPL '73})}. \bibinfo{publisher}{Association for
  Computing Machinery}, \bibinfo{address}{New York, NY, USA},
  \bibinfo{pages}{153–168}.
\newblock
\showISBNx{9781450373494}
\href{https://doi.org/10.1145/512927.512942}{doi:\nolinkurl{10.1145/512927.512942}}


\bibitem[Hintikka(1962)]%
        {hintikka1962knowledge}
\bibfield{author}{\bibinfo{person}{Jaakko Hintikka}.}
  \bibinfo{year}{1962}\natexlab{}.
\newblock \bibinfo{booktitle}{\emph{Knowledge and belief}}.
\newblock \bibinfo{publisher}{Cornell University Press}.
\newblock


\bibitem[Hitz et~al\mbox{.}(1994)]%
        {hitz1994file}
\bibfield{author}{\bibinfo{person}{Dave Hitz}, \bibinfo{person}{James Lau},
  {and} \bibinfo{person}{Michael~A Malcolm}.} \bibinfo{year}{1994}\natexlab{}.
\newblock \showarticletitle{{File System Design for an NFS File Server
  Appliance.}}. In \bibinfo{booktitle}{\emph{{USENIX Winter}}},
  Vol.~\bibinfo{volume}{94}.
\newblock


\bibitem[Jul et~al\mbox{.}(1988)]%
        {jul1988fine}
\bibfield{author}{\bibinfo{person}{Eric Jul}, \bibinfo{person}{Henry Levy},
  \bibinfo{person}{Norman Hutchinson}, {and} \bibinfo{person}{Andrew Black}.}
  \bibinfo{year}{1988}\natexlab{}.
\newblock \showarticletitle{Fine-grained mobility in the Emerald system}.
\newblock \bibinfo{journal}{\emph{ACM Transactions on Computer Systems (TOCS)}}
  \bibinfo{volume}{6}, \bibinfo{number}{1} (\bibinfo{year}{1988}),
  \bibinfo{pages}{109--133}.
\newblock


\bibitem[Jung et~al\mbox{.}(2018)]%
        {jung2018iris}
\bibfield{author}{\bibinfo{person}{Ralf Jung}, \bibinfo{person}{Robbert
  Krebbers}, \bibinfo{person}{Jacques-Henri Jourdan},
  \bibinfo{person}{Ale{\v{s}} Bizjak}, \bibinfo{person}{Lars Birkedal}, {and}
  \bibinfo{person}{Derek Dreyer}.} \bibinfo{year}{2018}\natexlab{}.
\newblock \showarticletitle{Iris from the ground up: A modular foundation for
  higher-order concurrent separation logic}.
\newblock \bibinfo{journal}{\emph{Journal of Functional Programming}}
  \bibinfo{volume}{28} (\bibinfo{year}{2018}), \bibinfo{pages}{e20}.
\newblock


\bibitem[Kamide(2002)]%
        {kamide2002kripke}
\bibfield{author}{\bibinfo{person}{Norihiro Kamide}.}
  \bibinfo{year}{2002}\natexlab{}.
\newblock \showarticletitle{Kripke semantics for modal substructural logics}.
\newblock \bibinfo{journal}{\emph{Journal of Logic, Language and Information}}
  \bibinfo{volume}{11}, \bibinfo{number}{4} (\bibinfo{year}{2002}),
  \bibinfo{pages}{453--470}.
\newblock


\bibitem[Krebbers et~al\mbox{.}(2018)]%
        {krebbers2018mosel}
\bibfield{author}{\bibinfo{person}{Robbert Krebbers},
  \bibinfo{person}{Jacques-Henri Jourdan}, \bibinfo{person}{Ralf Jung},
  \bibinfo{person}{Joseph Tassarotti}, \bibinfo{person}{Jan-Oliver Kaiser},
  \bibinfo{person}{Amin Timany}, \bibinfo{person}{Arthur Chargu{\'e}raud},
  {and} \bibinfo{person}{Derek Dreyer}.} \bibinfo{year}{2018}\natexlab{}.
\newblock \showarticletitle{MoSeL: A general, extensible modal framework for
  interactive proofs in separation logic}.
\newblock \bibinfo{journal}{\emph{Proceedings of the ACM on Programming
  Languages}} \bibinfo{volume}{2}, \bibinfo{number}{ICFP}
  (\bibinfo{year}{2018}), \bibinfo{pages}{1--30}.
\newblock


\bibitem[Krebbers et~al\mbox{.}(2017a)]%
        {krebbers2017essence}
\bibfield{author}{\bibinfo{person}{Robbert Krebbers}, \bibinfo{person}{Ralf
  Jung}, \bibinfo{person}{Ale{\v{s}} Bizjak}, \bibinfo{person}{Jacques-Henri
  Jourdan}, \bibinfo{person}{Derek Dreyer}, {and} \bibinfo{person}{Lars
  Birkedal}.} \bibinfo{year}{2017}\natexlab{a}.
\newblock \showarticletitle{The essence of higher-order concurrent separation
  logic}. In \bibinfo{booktitle}{\emph{European Symposium on Programming}}.
  Springer, \bibinfo{pages}{696--723}.
\newblock


\bibitem[Krebbers et~al\mbox{.}(2017b)]%
        {krebbers-17}
\bibfield{author}{\bibinfo{person}{Robert Krebbers}, \bibinfo{person}{Amin
  Timany}, {and} \bibinfo{person}{Lars Birkedal}.}
  \bibinfo{year}{2017}\natexlab{b}.
\newblock \showarticletitle{Interactive proofs in higher-order concurrent
  separation logic}. In \bibinfo{booktitle}{\emph{Principles of Programming
  Languages ({POPL})}}.
\newblock
\urldef\tempurl%
\url{http://cs.au.dk/~ birke/papers/ipm-conf.pdf}
\showURL{%
\tempurl}


\bibitem[Kuru and Gordon(2025)]%
        {kuru2025modalabstractionsvirtualizingmemory}
\bibfield{author}{\bibinfo{person}{Ismail Kuru} {and} \bibinfo{person}{Colin~S.
  Gordon}.} \bibinfo{year}{2025}\natexlab{}.
\newblock \showarticletitle{{Modal Abstractions for Virtualizing Memory
  Addresses}}.
\newblock \bibinfo{journal}{\emph{Proceedings of the ACM on Programming
  Languages}} \bibinfo{volume}{9}, \bibinfo{number}{OOPSLA2}
  (\bibinfo{year}{2025}).
\newblock
\href{https://doi.org/10.1145/3763134}{doi:\nolinkurl{10.1145/3763134}}


\bibitem[Lahav et~al\mbox{.}(2017)]%
        {lahav2017repairing}
\bibfield{author}{\bibinfo{person}{Ori Lahav}, \bibinfo{person}{Viktor
  Vafeiadis}, \bibinfo{person}{Jeehoon Kang}, \bibinfo{person}{Chung-Kil Hur},
  {and} \bibinfo{person}{Derek Dreyer}.} \bibinfo{year}{2017}\natexlab{}.
\newblock \showarticletitle{Repairing sequential consistency in C/C++ 11}. In
  \bibinfo{booktitle}{\emph{Proceedings of the 38th ACM SIGPLAN Conference on
  Programming Language Design and Implementation}}. \bibinfo{pages}{618--632}.
\newblock


\bibitem[Leino(2010)]%
        {leino2010dafny}
\bibfield{author}{\bibinfo{person}{K~Rustan~M Leino}.}
  \bibinfo{year}{2010}\natexlab{}.
\newblock \showarticletitle{Dafny: An automatic program verifier for functional
  correctness}. In \bibinfo{booktitle}{\emph{Logic for Programming, Artificial
  Intelligence, and Reasoning}}. Springer, \bibinfo{pages}{348--370}.
\newblock


\bibitem[Murphy et~al\mbox{.}(2004)]%
        {murphy2004symmetric}
\bibfield{author}{\bibinfo{person}{Tom Murphy}, \bibinfo{person}{Karl Crary},
  \bibinfo{person}{Robert Harper}, {and} \bibinfo{person}{Frank Pfenning}.}
  \bibinfo{year}{2004}\natexlab{}.
\newblock \showarticletitle{A symmetric modal lambda calculus for distributed
  computing}. In \bibinfo{booktitle}{\emph{Proceedings of the 19th Annual IEEE
  Symposium on Logic in Computer Science, 2004.}} IEEE,
  \bibinfo{pages}{286--295}.
\newblock


\bibitem[Nakano(2000)]%
        {nakano-00}
\bibfield{author}{\bibinfo{person}{Hiroshi Nakano}.}
  \bibinfo{year}{2000}\natexlab{}.
\newblock \showarticletitle{A Modality for Recursion}. In
  \bibinfo{booktitle}{\emph{Logic in Computer Science (LICS)}}.
  \bibinfo{pages}{255--266}.
\newblock
\urldef\tempurl%
\url{http://www602.math.ryukoku.ac.jp/~ nakano/papers/modality-lics00.ps.gz}
\showURL{%
\tempurl}


\bibitem[O'Hearn and Pym(1999)]%
        {ohearn1999bunched}
\bibfield{author}{\bibinfo{person}{Peter~W O'Hearn} {and}
  \bibinfo{person}{David~J Pym}.} \bibinfo{year}{1999}\natexlab{}.
\newblock \showarticletitle{The logic of bunched implications}.
\newblock \bibinfo{journal}{\emph{Bulletin of Symbolic Logic}}
  \bibinfo{volume}{5}, \bibinfo{number}{2} (\bibinfo{year}{1999}),
  \bibinfo{pages}{215--244}.
\newblock


\bibitem[Owens et~al\mbox{.}(2009)]%
        {owens2009better}
\bibfield{author}{\bibinfo{person}{Scott Owens}, \bibinfo{person}{Susmit
  Sarkar}, {and} \bibinfo{person}{Peter Sewell}.}
  \bibinfo{year}{2009}\natexlab{}.
\newblock \showarticletitle{A better x86 memory model: x86-TSO}.
\newblock In \bibinfo{booktitle}{\emph{Theorem Proving in Higher Order
  Logics}}. \bibinfo{publisher}{Springer}, \bibinfo{pages}{391--407}.
\newblock


\bibitem[Pnueli(1977)]%
        {pnueli1977temporal}
\bibfield{author}{\bibinfo{person}{Amir Pnueli}.}
  \bibinfo{year}{1977}\natexlab{}.
\newblock \showarticletitle{The temporal logic of programs}. In
  \bibinfo{booktitle}{\emph{18th Annual Symposium on Foundations of Computer
  Science (FOCS 1977)}}. ieee, \bibinfo{pages}{46--57}.
\newblock


\bibitem[Pratt(1976)]%
        {pratt1976semantical}
\bibfield{author}{\bibinfo{person}{Vaughan~R Pratt}.}
  \bibinfo{year}{1976}\natexlab{}.
\newblock \showarticletitle{Semantical consideration on {Floyd-Hoare} logic}.
  In \bibinfo{booktitle}{\emph{Foundations of Computer Science, 1976., 17th
  Annual Symposium on}}. IEEE, \bibinfo{pages}{109--121}.
\newblock


\bibitem[Restall(1993)]%
        {restall1993modalities}
\bibfield{author}{\bibinfo{person}{Greg Restall}.}
  \bibinfo{year}{1993}\natexlab{}.
\newblock \showarticletitle{Modalities in substructural logics}.
\newblock \bibinfo{journal}{\emph{Logique et Analyse}} \bibinfo{volume}{36},
  \bibinfo{number}{141/142} (\bibinfo{year}{1993}), \bibinfo{pages}{25--38}.
\newblock


\bibitem[Restall(2002)]%
        {restall2002introduction}
\bibfield{author}{\bibinfo{person}{Greg Restall}.}
  \bibinfo{year}{2002}\natexlab{}.
\newblock \bibinfo{booktitle}{\emph{An introduction to substructural logics}}.
\newblock \bibinfo{publisher}{Routledge}.
\newblock


\bibitem[Rodeh et~al\mbox{.}(2013)]%
        {rodeh2013btrfs}
\bibfield{author}{\bibinfo{person}{Ohad Rodeh}, \bibinfo{person}{Josef Bacik},
  {and} \bibinfo{person}{Chris Mason}.} \bibinfo{year}{2013}\natexlab{}.
\newblock \showarticletitle{{BTRFS: The Linux B-Tree Filesystem}}.
\newblock \bibinfo{journal}{\emph{ACM Trans. Storage}} \bibinfo{volume}{9},
  \bibinfo{number}{3}, Article \bibinfo{articleno}{9} (\bibinfo{date}{Aug.}
  \bibinfo{year}{2013}), \bibinfo{numpages}{32}~pages.
\newblock
\showISSN{1553-3077}
\href{https://doi.org/10.1145/2501620.2501623}{doi:\nolinkurl{10.1145/2501620.2501623}}


\bibitem[Rosenblum and Ousterhout(1992)]%
        {Rosenblum1992LFS}
\bibfield{author}{\bibinfo{person}{Mendel Rosenblum} {and}
  \bibinfo{person}{John~K. Ousterhout}.} \bibinfo{year}{1992}\natexlab{}.
\newblock \showarticletitle{The Design and Implementation of a Log-structured
  File System}.
\newblock \bibinfo{journal}{\emph{ACM Trans. Comput. Syst.}}
  \bibinfo{volume}{10}, \bibinfo{number}{1} (\bibinfo{date}{Feb.}
  \bibinfo{year}{1992}), \bibinfo{pages}{26--52}.
\newblock
\showISSN{0734-2071}
\href{https://doi.org/10.1145/146941.146943}{doi:\nolinkurl{10.1145/146941.146943}}


\bibitem[Sieczkowski et~al\mbox{.}(2015)]%
        {sieczkowski2015separation}
\bibfield{author}{\bibinfo{person}{Filip Sieczkowski}, \bibinfo{person}{Kasper
  Svendsen}, \bibinfo{person}{Lars Birkedal}, {and} \bibinfo{person}{Jean
  Pichon-Pharabod}.} \bibinfo{year}{2015}\natexlab{}.
\newblock \showarticletitle{{A separation Logic for Fictional Sequential
  Consistency}}.
\newblock In \bibinfo{booktitle}{\emph{European Symposium on Programming}}.
  \bibinfo{publisher}{Springer}, \bibinfo{pages}{736--761}.
\newblock


\bibitem[Strom and Yemini(1986)]%
        {strom1986typestate}
\bibfield{author}{\bibinfo{person}{Robert~E Strom} {and}
  \bibinfo{person}{Shaula Yemini}.} \bibinfo{year}{1986}\natexlab{}.
\newblock \showarticletitle{Typestate: A programming language concept for
  enhancing software reliability}.
\newblock \bibinfo{journal}{\emph{IEEE transactions on software engineering}}
  \bibinfo{number}{1} (\bibinfo{year}{1986}), \bibinfo{pages}{157--171}.
\newblock


\bibitem[Vindum et~al\mbox{.}(2025)]%
        {larsnextgen25}
\bibfield{author}{\bibinfo{person}{Simon~Friis Vindum},
  \bibinfo{person}{A\"{\i}na~Linn Georges}, {and} \bibinfo{person}{Lars
  Birkedal}.} \bibinfo{year}{2025}\natexlab{}.
\newblock \showarticletitle{The Nextgen Modality: A Modality for
  Non-Frame-Preserving Updates in Separation Logic}. In
  \bibinfo{booktitle}{\emph{Proceedings of the 14th ACM SIGPLAN International
  Conference on Certified Programs and Proofs}} (Denver, CO, USA)
  \emph{(\bibinfo{series}{CPP '25})}. \bibinfo{pages}{83–97}.
\newblock
\showISBNx{9798400713477}
\href{https://doi.org/10.1145/3703595.3705876}{doi:\nolinkurl{10.1145/3703595.3705876}}


\bibitem[Wagner et~al\mbox{.}(2024)]%
        {amalreal2024}
\bibfield{author}{\bibinfo{person}{Andrew Wagner}, \bibinfo{person}{Zachary
  Eisbach}, {and} \bibinfo{person}{Amal Ahmed}.}
  \bibinfo{year}{2024}\natexlab{}.
\newblock \showarticletitle{Realistic Realizability: Specifying {ABI}s You Can
  Count On}.
\newblock \bibinfo{journal}{\emph{Proc. ACM Program. Lang.}}
  \bibinfo{volume}{8}, \bibinfo{number}{OOPSLA2}, Article
  \bibinfo{articleno}{315} (\bibinfo{date}{Oct.} \bibinfo{year}{2024}),
  \bibinfo{numpages}{30}~pages.
\newblock
\href{https://doi.org/10.1145/3689755}{doi:\nolinkurl{10.1145/3689755}}


\end{thebibliography}
\appendix
\end{document}